\documentclass{pic2012}
\usepackage{cite}

\graphicspath{{pdf/}{jpeg/}{png/}{plots/}{./}}
\DeclareGraphicsExtensions{.pdf,.jpeg,.png}

\usepackage[tight,footnotesize]{subfigure}

\usepackage{url}

\def\runauthor{I.~Marfin}
\def\shorttitle{Search for Higgs boson in association with b quarks}

\begin{document}

\title{Search for Higgs boson production in association with b quarks at CMS in pp collisions}

\author{ 
 Igor~Marfin  on behalf of CMS
}


\address{ DESY, Notkestrasse 85, 22607 Hamburg, Germany \\  BTU, Platz de Deutschen Einheit 1, 03046 Cottbus, Germany}

\maketitle

\abstracts{
A search for neutral Higgs bosons produced in association with b-quark(s) and decaying into a pair of
b-quarks is performed with the CMS  detector at LHC.
The Higgs boson signal is expected to emerge as an excess in the mass spectrum of
two b-tagged jets.
Dedicated triggers with on-line b-tagging in fully hadronic events were developed specifically
for this kind of analysis. Limits on the cross section times branching fractions are derived model independently.
The result was interpreted in the Minimal Supersymmetric Model (MSSM). Upper limits at $tan\beta$ are derived as
a function of $M_A$, the mass of the Higgs boson $A$.
In the analysis presented here data taken in 2011 at 7 TeV are used
corresponding to a total luminosity of $L=2.7 -– 4.0 fb^{-1}$ for low and medium Higgs boson masses, respectively.
}

\section{Introduction}

A search is performed for a Higgs boson produced in association with additional b-quarks. 
In the Standard Model (SM) the cross section 
is very low and due to the large
background dominated by QCD processes
 with three b-quark or two b-quark plus c,udsg-quark final states the 
signal  in the current data is not in reach.

In the Minimal Super-symmetric Model (MSSM)\cite{Nilles:1983ge,Haber:1984rc}, there are 3 neutral Higgs bosons, one  CP-odd 
($A$) and two CP-even ($h,H$). Their couplings 
to d-type fermions are enhanced by a factor $tan\beta$. The Higgs bosons $A$ and $h(H)$ degenerate in the mass. Hence,
the cross section of Higgs boson production in association with b quarks
increases by a factor $2\times tan^{2}\beta$.

The MSSM Higgs bosons production is  dominated at hadron colliders by the sub-process $gg\rightarrow b(\bar{b})\Phi \,+\, X$ with
only a small contribution from $\bar{q}q\rightarrow \bar{b}b\Phi \,+\, X$.
The  branching fraction of the Higgs boson decay to b-quarks are in the range
 $[70\%, 92\%]$ \cite{Heinemeyer:2004ms}, and it remains large even at large masses of the Higgs boson $A$.

The dominant background arises from heavy flavor multi-jet QCD processes.
Other background contributions  like $Zbb$, $Wbb$, $t\bar{t}$ have low rates.

Data collected with the CMS detector\cite{CMS} 
in proton-proton collisions at a centre-of-mass energy of 7 TeV at the LHC,
corresponding to an integrated luminosity of $2.7-4.0 fb^{-1}$, were used in this search.
Specialized triggers have been developed, using
algorithms to tag b-quark jets from displaced vertices of B-meson decays.

\section{Trigger and event selection}

At LHC multi-jet events are produced copiously via higher order QCD processes.
To trigger events with jets originating from the decay of heavy particles
thresholds on  transverse momenta ($p_T$) of the jets are imposed in the first level trigger. 
In the Higgs boson low-mass search, performed in the  
data taken in the first period, the thresholds were $46\, GeV/c$ for the leading jet and $38\, GeV/c$ for 
the second leading jet.
In the second half of 2011, due to the higher luminosity, the threshold for the leading jet was  $60\,GeV/c$,
and for the second leading jet  $53\,GeV/c$. This data were used in the search for the medium-mass 
Higgs boson.
On-line b-tagging was performed with TCHE  algorithm \cite{Btag:TCHE} using the track 
with $2^{nd}$ highest significance, and at least two jets must be b-tagged.  
Tracks were clustered into on-line jets \cite{Cacciari:2008gp} with a cone  radius  $R=0.5.$

The jets were reconstructed using the particle-flow algorithm~\cite{CMS:PF,CMS:PF2,CMS:PF3} and
a standard jet energy calibration was applied~\cite{CMS:JET2}. Then,
jet energies were corrected for the contributions of pile-up (PU) interactions~\cite{CMS:JET1}. 
The combined secondary vertex (CSV) algorithm \cite{CSV1} was applied
for the offline identification of b-jets.
Secondary vertices (SV)  were reconstructed using the Adaptive Vertex Finder \cite{SV1}.
It combines several topological and kinematic variables of the SV and track impact parameters 
into a ratio of likelihoods. 
The ratio  was required to be larger or equal to $0.89$ (tight operation point, CSVT) to tag the jet.
Mistagged light flavor jets are suppressed to a level of $\sim 0.2\%$ with a b-jet 
identification efficiency of about $55\%$~\cite{BTAG}. 

A variable (EventBta) was created exploiting the SV mass spectra obtained for jets of different flavors.
EventBtag combines the SV masses of the three leading jets in one quantity which is small for c- and light-jets 
and large for b-jets.  

The three leading jets 
have  to satisfy the following criteria in the search of the  Higgs bosons with low (medium) masses:
$ p_{T}(jet_1)\geq 46(60)$, $p_T(jet_2)\geq 38(53)$, $p_T(jet_3)\geq 20$ GeV/c; $|\eta(jets)|<2.2$;
 at least three b-tagged jets with CSVT are required.

\section{Background Modeling}  

The signature of the process $pp\rightarrow bH\,+\, X\rightarrow 3b\, +\, X$ 
is at least three b-jets which are identified with the b-tagging algorithm. 
The dominant background is due to multijet QCD processes mainly composed of events with three b-jets
and events with two b-jets plus a charm- or udsg(light)-mistagged jet. 
Contributions from other processes like $t\bar{t}$, Z+jets 
were estimated to be negligible.

\begin{figure}[!t]
\centering
\subfigure[ ]{
\includegraphics[width=2.3in]{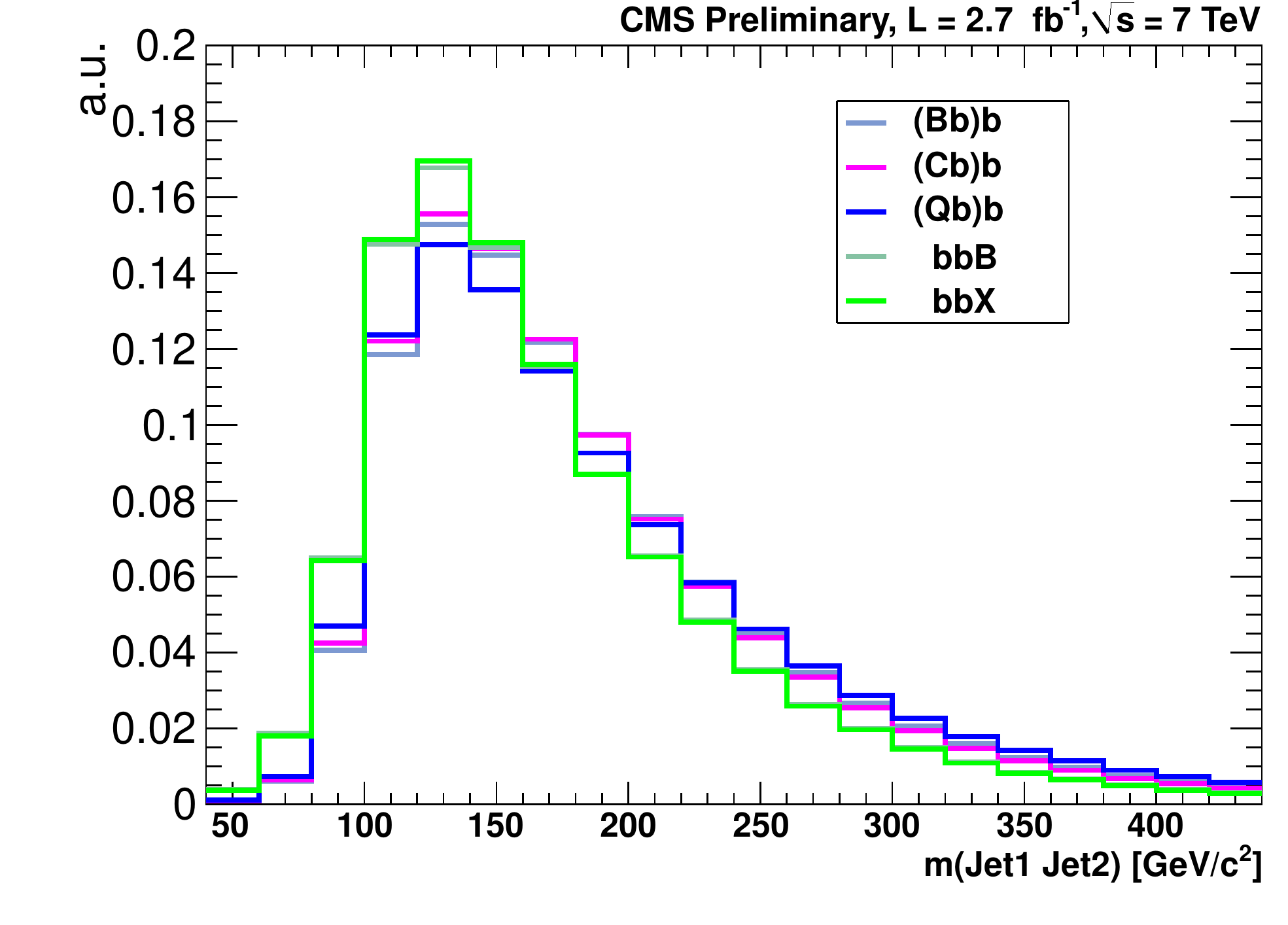}
\label{fig:a}
}
\subfigure[ ]{
\includegraphics[width=2.3in]{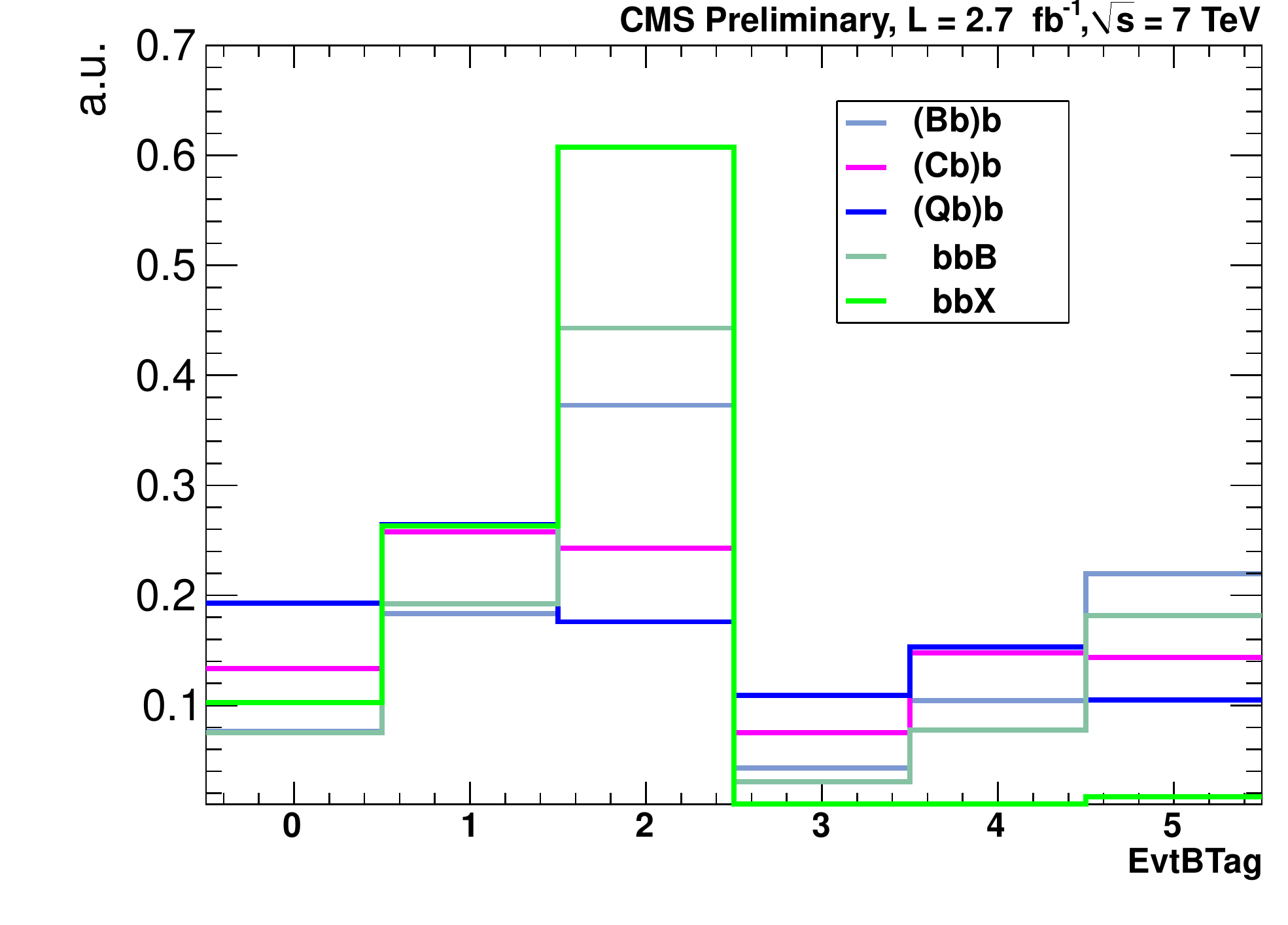}
\label{fig:b}
}
\caption{ $M_{12}$ (a) and EventBTag (b)  projections for the five background templates used
in the search for a low-mass Higgs boson. }
\label{fig:templates}
\end{figure}

We have used a data sample with two b-tagged jets 
to estimate the background in data with three b-tagged jets. The double-btag data events 
were properly weighted assuming the flavor (b,c,udsg) of the not-tagged jet  to obtain the shape of QCD background.
Then a 2D template 
spanned by the di-jet mass of first two leading jets, $M_{12}$  and EventBtag was created  by filling  
 with the events of the given category. 
Finally, five templates were used. Their projections in $M_{12}$ and EventBtag are shown in 
Figure \ref{fig:templates}.
Two additional corrections are applied  to remove contamination from non-bb events in the double-btag sample
and to account for differences between offline and on-line b-tag efficiencies \cite{our:note}.

Signal templates were obtained for each mass point from samples of Higgs bosons events simulated using $PYTHIA$ \cite{PYTHIA}.
Pileup reweighting and Level-1/Level-2 trigger efficiencies \cite{our:note} were applied.


\section{Results}

The signal and background yields were obtained from a binned unconstrained least-squares fit of '2D' templates to 
the triple-btag data distributions 
in the $M_{12}$ and EventBtag space \cite{our:note}. 
The statistical uncertainties of 
the templates are  taken into account.  

\begin{figure}[!t]
\centering
\subfigure[ ]{
\includegraphics[width=2.3in]{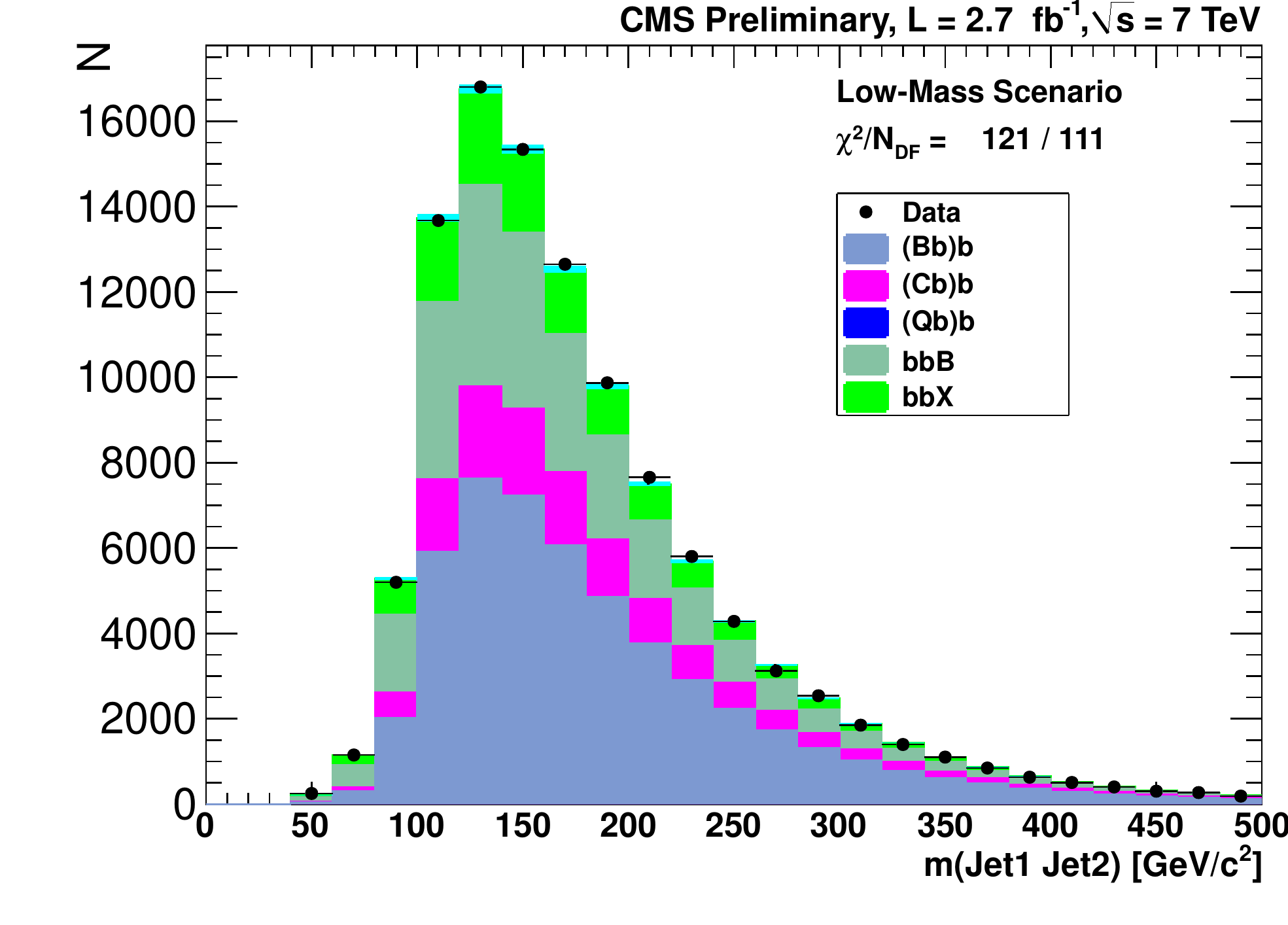}
}\label{fig:bkgonlyM12}
\subfigure[ ]{
\includegraphics[width=2.3in]{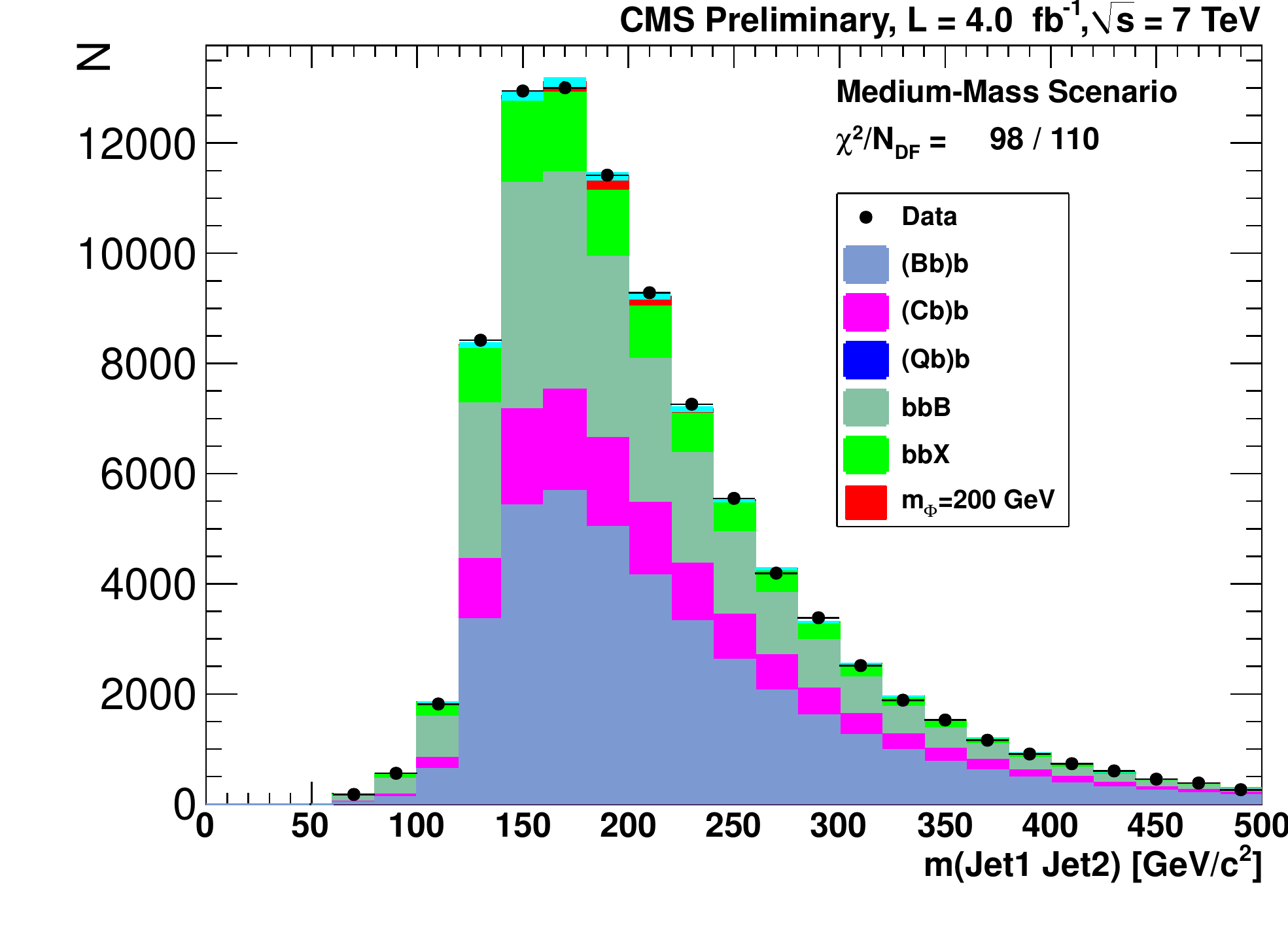}
\label{fig:bkgonlyEvtBtag}
}

\caption{
(a) Results of the 'background-only' fit in the triple-btag sample in the projection on the di-jet mass, 
(b) Results of the fit in the di-jet mass projection including a signal template for a Higgs boson with a mass of $200\, GeV/c^2$
in the triple-btag data for the medium-mass scenario.
}\label{fig:bkgonly}
\end{figure}

The fit to the data distribution was performed using only background templates. 
The result for the low-mass scenario of  the Higgs bosons search is shown in Figure \ref{fig:bkgonly}(a). 

The data distribution is well described by the fit, with a $\chi^2/N_{dof}\sim 1$.  
The data points
are within the uncertainty of the fit, indicated by the cyan area.
The same 'background-only-fit'  for the medium-mass scenario  also perfectly describes the
data distribution.
In both cases events with at least three b jets 
constitute the largest contribution to to the background. For the low-mass Higgs bosons search
it amounts to about $74\%$, in good agreement with the MC background estimation  of $72\%$.

\begin{figure}[!t]
\centering
\subfigure[ ]{
\includegraphics[width=2.3in]{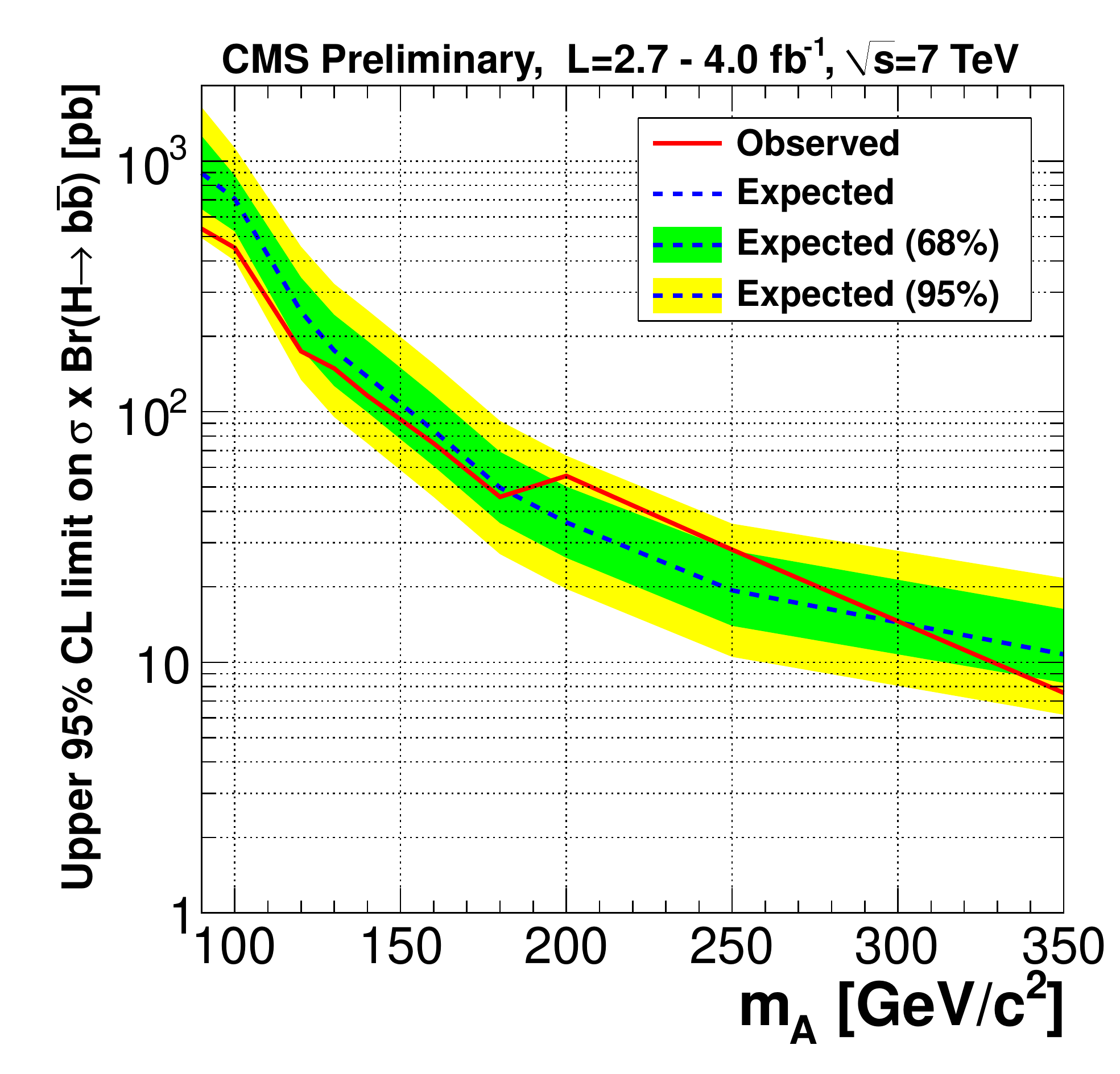}
} 
\subfigure[ ]{
\includegraphics[width=2.3in]{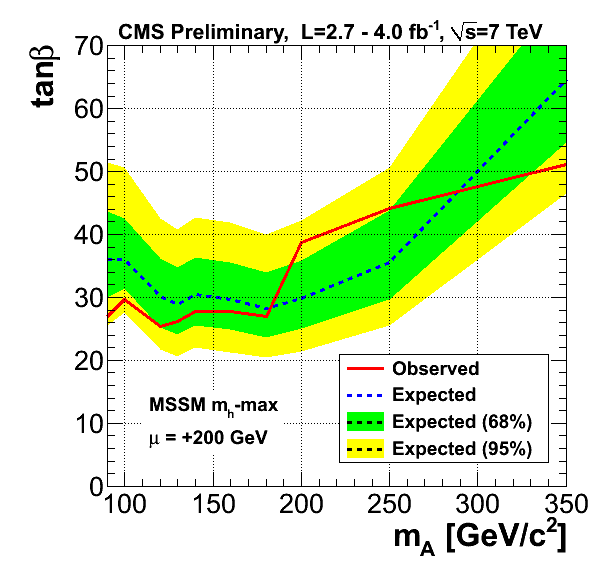}
} 
\caption{
 (a) The observed $95\%\,\, C.L.$ upper limits  on the cross section multiplied by the branching fraction as a function of $M_A$.
 (b) Observed and median expected $95\%$ C.L. upper limits on $tan\beta$  as a function of $M_A$
in the $m_h^{max}$ scenario for $\mu=+200$ $GeV/c^2$  
}\label{fig:upperlimits}
\end{figure}

The fits were repeated including signal templates for several Higgs boson masses
between $90$ to $350$ $GeV/c^2$.
As an example, the result of the fit assuming a Higgs boson with a mass of $200$ $GeV/c^2$ is shown in
Figure \ref{fig:bkgonly}(b). 
The values of the signal fractions obtained in the fits using all Higgs boson mass hypotheses
are used to derive signal cross sections multiplied with the branching fraction of the Higgs boson in a
pair of b-quarks
as shown in Figure \ref{fig:upperlimits}(a).


Upper limits on the cross section multiplied by the branching fraction,
$\sigma(pp\rightarrow b\Phi)\times BR(\Phi\rightarrow b\bar{b})$,
were calculated in the $[90,350]\,GeV/c^2$  range for $M_A$.
The modified frequentist $CLs$
method \cite{CLS1} is applied using the $RooStats$ package~\cite{ROOSTATS}.
As test statistics for the $CLs$ method the
profile likelihood ratio distribution was used.
Systematic uncertainties were taken into account as nuisance parameters of the likelihood fit to the data.
Normalizations of the templates were unconstrained in the fit. 
The observed  and the median expected $95\%\,\, C.L.$ limits as functions of the Higgs boson mass are
shown in  Figure \ref{fig:upperlimits}(b). The observed limit is well within the
expected $2\sigma$ band.

The model independent limits on  $\sigma(pp\rightarrow b\Phi)\times BR(\Phi\rightarrow b\bar{b})$
were translated into 
$tan\beta$ limits within the $m_h^{max}$ scenario~\cite{MHMAX}, where the parameter
 $\mu$ is set to $+ 200 $ and $-200 GeV/c^2$. 

The expected cross  section and branching fraction, in the MSSM framework, 
are calculated by $BBH@NNLO$~\cite{BBHNNLO} and
$FEYNHIGGS$~\cite{FEYNHIGGS1}-\cite{FEYNHIGGS4}, respectively.

The observed  and expected median  $95\%\,\, C.L.$ upper limits on $tan\beta$
as a function of $M_A$  using
$\mu=+200 GeV/c^2$ are shown in Figure \ref{fig:upperlimits}(b).
The upper limits on $tan\beta$ 
for $\mu=-200$ $GeV/c^2$  case were compared with the recent
results from D0~\cite{D0} and CDF~\cite{CDF} experiments at the Tevatron collider.
The limits obtained in the current analysis are  significantly lower than
the Tevatron results.


\end{document}